\documentclass{elsarticle}
\usepackage{lineno,hyperref}

\usepackage{esvect}
\usepackage{latexsym}
\usepackage{graphicx}
\usepackage{amsmath}

\journal{Nuclear Instruments and Methods}

\begin{document}
\begin{frontmatter}

\title{Extracting  the depolarization coefficient $D_{NN}$ from data measured with a full acceptance detector}
\author[1,3]   { F.~Hauenstein}
\author[4,5]   { H.~Clement}
\author[1]     { R.~Dzhygadlo\fnref{f1}}
\author[3]     { W.~Eyrich}
\author[1]     { A.~Gillitzer}
\author[1]     { D.~Grzonka}
\author[1,6]   { S.~Jowzaee}
\author[1,2,7] { J.~Ritman}
\author[1]     { E.~Roderburg\fnref{f2}}
\author[1]     { M.~R\"{o}der\fnref{f3}}
\author[1]     { P.~Wintz}                  

\address[1] {Institut f\"{u}r Kernphysik, Forschungszentrum J\"{u}lich, 52428 J\"{u}lich, Germany}
\address[2] {J\"{u}lich Aachen Research Allianz, Forces and Matter Experiments (JARA-FAME)}
\address[3] {Friedrich-Alexander-Universit\"{a}t Erlangen-N\"{u}rnberg, 91058 Erlangen, Germany} 
\address[4] {Physikalisches Institut der Universit\"{a}t T\"{u}bingen, Auf der Morgenstelle 14, 72076 T\"{u}bingen, Germany} 
\address[5] {Kepler Center for Astro and Particle Physics, University of T\"{u}bingen, Auf der Morgenstelle 14, 72076 T\"{u}bingen, Germany}
\address[6] {Institute of Physics, Jagiellonian University, 30-059 Cracow Poland} 
\address[7] {Experimentalphysik I, Ruhr-Universit\"{a}t Bochum, 44780 Bochum, Germany}
\fntext[f1] {current address: Hadron Physics I, GSI Helmholtzzentrum f\"{u}r Schwerionenforschung GmbH}
\fntext[f2] {corresponding author e.roderburg@fz-juelich.de}
\fntext[f3] {current address: Corporate Development, Forschungszentrum J\"{u}lich, 52428 J\"{u}lich, Germany}

\begin {abstract}
The spin transfer from  vertically polarized beam protons to 
$\Lambda$ or $\Sigma$ hyperons of the associated strangeness
production $\vv{\mathrm{p}}$p $\rightarrow$ p$\mathrm{K}^{+}\Lambda$ 
and $\vv{\mathrm{p}} $p $\rightarrow$ p$\mathrm{K}^{0}\Sigma^{+}$ is described
with the depolarization coefficient $D_{NN}$.  As the polarization of
the hyperons is determined by their weak decays, detectors, which have
a large acceptance for the decay particles, are needed. In this paper
a formula is derived, which describes the depolarization coefficient
$D_{NN}$ by count rates of a 4$\pi$ detector.  It is shown, that
formulas, which are given in publications for detectors with
restricted acceptance, are specific cases of this formula for a 4$\pi$
detector.
\end{abstract}
\end{frontmatter}
\section{Introduction }
The determination of the depolarization coefficient of the
asso\-ciated strange\-ness production reaction is an important tool to
gain information about the quark configuration of the initial proton
and the produced hyperon. (see e.g. \cite {Kubo1999} and references
therein).  While analyzing the polarization data measured with the
time of flight spectrometer (COSY-TOF) (see e.g. \cite
{Roeder2013}) it was noticed that formulas given in literature for
the depolarization coefficient $D_{NN}$
are tailored to the acceptances of the respective detectors 
\cite {Bonner1988,Bravar1997,Balestra1999,Balestra1999a,Maggiora2001}. Therefore,
they are not valid for the COSY-TOF experiment, which is a 4$\pi$ acceptance detector.
In this paper a formula for a 4$\pi$ acceptance detector is derived.
 This formula can be applied for the reactions $\vv{\mathrm{p}}$p $\rightarrow$ p$\mathrm{K}^{+}\Lambda$ 
and $\vv{\mathrm{p}}$p $\rightarrow$ p$\mathrm{K}^{0}\Sigma^{+}$. For the intermediate $\Sigma^{0}$ production in 
$\vv{\mathrm{p}} $p $\rightarrow$ p$\mathrm{K}^{+}\Sigma^{0}$ $\rightarrow$  p$\mathrm{K}^{+}\Lambda \gamma$ 
the $D_{NN}$ formula can not be applied, as in this reaction the $\Lambda$ polarization is diluted by the $\Sigma^{0}$
polarization and by the additional degree of freedom introduced by the photon momentum \cite {Gatto1957}.

Measurements of $D_{NN}$ with limited acceptance detectors can be described by modifying this
formula for the respective acceptance. This is demonstrated with the formula given for the
Brookhaven Multiparticle Spectrometer \cite {Bonner1988}.

The analyzing power and  the hyperon
polarization are discussed first, in order to introduce the ``Integral
Method'' and to define some of the variables.
Here and in the
  following we apply the notations of E. B. Bonner et al.  \cite {Bonner1988}, which are in
  agreement with the Ann Arbor Conventions \cite{Ashkin1978}.
\section{Analyzing power }

The hyperon analyzing power $A_N$ describes how strong the
(transversely) polarized beam generates a left right asymmetry of the
hyperon. It can be measured by applying the formula:

\begin{equation}\label{eq:analyzingpower}
I(\vartheta^*,\Phi) =  I_0(\vartheta^*)\cdot(1+A_N(\vartheta^*) P_B \cos(\Phi))
\end{equation}

\begin{figure}[htb]
	\centering
	\includegraphics[width=.8\textwidth]{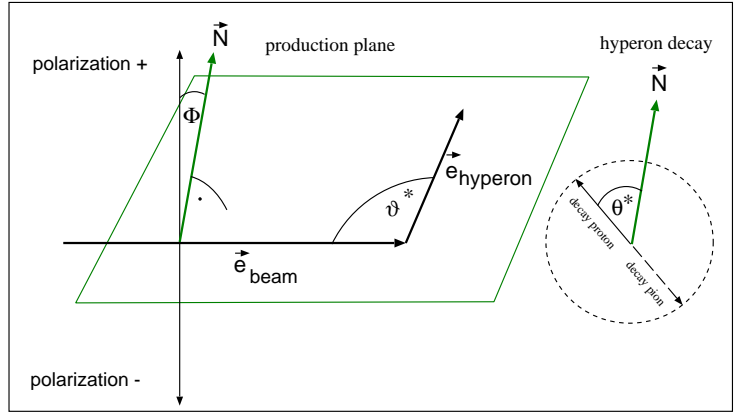}
	\caption{\label{definitions} Schematic illustration of the
          normal vector $\vec{\mathrm{N}}$ to the production plane, 
          which is defined as the cross product of $\vec{e}_{\mathrm{beam}}$ and
          $\vec{e}_{\mathrm{hyperon}}$,
          the angle $\Phi$ between $\vec{\mathrm{N}}$ and the
          direction of the beam polarization, and the angle
          $\theta^{*}$ of the decay nucleon in the hyperon rest system
          to $\vec{\mathrm{N}}$.  
        }
\end{figure}

$I$ is the intensity distribution, $ P_B$ the absolute value of the
beam polarization, and $\Phi$ is the angle between the direction of
the beam polarization ($\pm$ $\vv{e_y}$) and the normal vector $\vv{\mathrm{N}}$ to the
production plane ($\vv{e}_{beam}$ x $\vv{e}_{hyperon}$)
(s. fig. \ref{definitions}). $\vartheta^*$ is the polar scattering
angle in the cm system.  Alternatively, the polarization variables are
often evaluated as a function of the transverse momentum, Feynman x,
or the rapidity of the hyperon.

For the beam polarization pointing upwards (+ $\vv{e_y}$) the
azimuthal angle of the hyperon $\varphi$ is connected to the angle
$\Phi$ :
\newline $\varphi$ =  $\Phi$

For the beam polarization pointing downwards (- $\vv{e_y}$):
\newline $\varphi$ =  $\Phi$ - $\pi$

In the following the ``Integral Method'' of the evaluation of the
analyzing power is given. Dependent on the position in the left or
right hemisphere of the hyperon azimuthal angle $\varphi$ and on the
direction of the beam polarization, four count rates can be
distinguished:

$\varphi$ in the left hemisphere and beam polarization upwards

\begin{equation}N_L^{\uparrow}(\vartheta^*) = \int_0^1 I(\vartheta^*,\Phi)\, \mathrm{d}\!\cos(\Phi)    \end{equation}

$\varphi$ in the right hemisphere and beam polarization upwards   

\begin{equation}N_R^{\uparrow}(\vartheta^*) = \int_{-1}^0 I(\vartheta^*,\Phi) \, \mathrm{d}\!\cos(\Phi)  \end{equation}  

$\varphi$ in the left hemisphere and beam polarization downwards  

\begin{equation}N_L^{\downarrow}(\vartheta^*) = \int_{-1}^0 I(\vartheta^*,\Phi) \, \mathrm{d}\!\cos(\Phi)  \end{equation}

$\varphi$ in the right hemisphere and beam polarization downwards  

\begin{equation}N_R^{\downarrow}(\vartheta^*) = \int_0^1 I(\vartheta^*,\Phi) \, \mathrm{d}\!\cos(\Phi)    \end{equation}    

Assuming that the absolute value of the beam polarization is the same for both directions and that the detector is fully symmetric in $\varphi$ 
the count rates $N_L^{\uparrow}(\vartheta^*)$, $N_R^{\downarrow}(\vartheta^*)$ 
and $N_R^{\uparrow}(\vartheta^*)$, $N_L^{\downarrow}(\vartheta^*)$ are pairwise the same.  

The integration of equation \ref{eq:analyzingpower} yields:

\begin{equation}\label{eq:analyzingpower2}
A_N(\vartheta^*) = \frac{2} {P_B} \epsilon_A(\vartheta^*)
\end{equation}

with the averaged asymmetry

\begin{equation}\label{eq:analyzingpower3}
 \epsilon_A(\vartheta^*) = \frac{(N_L^{\uparrow}(\vartheta^*)+N_R^{\downarrow}(\vartheta^*))-(N_R^{\uparrow}(\vartheta^*)+N_L^{\downarrow}(\vartheta^*))}{N_L^{\uparrow}(\vartheta^*)+N_R^{\downarrow}(\vartheta^*)+N_R^{\uparrow}(\vartheta^*)+N_L^{\downarrow}(\vartheta^*)}
\end{equation}

\section{Hyperon  polarization }

It has been discovered that the hyperons are produced polarized even with
an unpolarized beam \cite {Lesnik1975}. The polarization axis is
normal to the production plane. It is determined by measuring
the angle $\theta^*$ between the direction of the daughter baryon (in the hyperon
rest frame) and the normal vector to the production plane $\vv{\mathrm{N}}$
 (s. fig. \ref{definitions}). The hyperon polarization $P_N$
can be determined by the equation:

 \begin{equation}\label{eq:hyperonpolarization}
I(\vartheta^*,\theta^*) =  I_0(\vartheta^*)\cdot(1+P_N(\vartheta^*)\: \alpha \cos(\theta^*))
\end{equation}


with the hyperon decay asymmetry parameter $\alpha$
\newline
($\alpha(\Lambda\rightarrow p \pi^{-}) = 0.642 \pm 0.013 $,  $\alpha(\Sigma^+ \rightarrow p \pi^0) = - 0.980 + 0.017 - 0.052$ \cite {Olive2014} ).

Several methods of evaluating $P_N(\vartheta^*)$ from the data can be applied: The count rate distribution versus $\cos(\theta^*)$ can be
fitted with a first order polynomial, the inclination represents $P_N(\vartheta^*)\: \alpha $. For this method the measured distribution
has to be normalized to the distribution derived from Monte Carlo simulations. Another method is the calculation of the weighted sum:
$P_N(\vartheta^*)\: \alpha = \sum{\cos(\theta^*)}/\sum{\cos^{2}(\theta^*)}$,  which was developed in \cite {Besset1979b} 

Here we apply the integral method, in order
to prepare the variables for the derivation of the spin transfer equation.
Depending on whether the  hyperon decay particle is emitted above (same hemisphere as $\vv{\mathrm{N}}$) or
 below (opposite hemisphere as  $\vv{\mathrm{N}}$) the production plane and on the azimuthal angle of the hyperon $\varphi$ (left or right
hemisphere) four count rates can be defined:

\noindent

 $\varphi$ in the left hemisphere and the decay particle above the production plane:

\begin{equation}N_L^A(\vartheta^*) = \int_0^1 I(\vartheta^*,\theta^*) \, \mathrm{d}\!\cos(\theta^*)     \end{equation}

 $\varphi$ in the right hemisphere and the decay particle above the production plane:

\begin{equation}N_R^A(\vartheta^*) = \int_0^1 I(\vartheta^*,\theta^*) \, \mathrm{d}\!\cos(\theta^*)     \end{equation}

 $\varphi$ in the left hemisphere and the decay particle below the production plane:

\begin{equation}N_L^B(\vartheta^*) = \int_{-1}^0 I(\vartheta^*,\theta^*) \, \mathrm{d}\!\cos(\theta^*)   \end{equation}

 $\varphi$ in the right hemisphere and the decay particle below the production plane:

\begin{equation}N_R^B(\vartheta^*) = \int_{-1}^0 I(\vartheta^*,\theta^*) \, \mathrm{d}\!\cos(\theta^*)   \end{equation}

$N_L^A(\vartheta^*)$ and $ N_R^B(\vartheta^*)$ correspond to the hyperon polarization in  (+ $\vv{e_y}$) direction. They are 
identical because the cross section is symmetric in $\varphi$, an unpolarized beam is assumed and detector asymmetries are ignored.
 The same holds for $N_L^B(\vartheta^*)$ and $ N_R^A(\vartheta^*)$ for 
the hyperon polarization in the  (- $\vv{e_y}$) direction.

The integration of equation \ref{eq:hyperonpolarization} yields:

\begin{equation}\label{eq:hyperonpolarization2}
P_N(\vartheta^*) = \frac{2} {\alpha} \epsilon_P(\vartheta^*)
\end{equation}

with the asymmetry

\begin{equation}\label{eq:hyperonpolarization3}
 \epsilon_P(\vartheta^*) = \frac{(N_L^A(\vartheta^*)+N_R^B(\vartheta^*))-(N_L^B(\vartheta^*)+N_R^A(\vartheta^*))}
                                { N_L^A(\vartheta^*)+N_R^B(\vartheta^*) + N_L^B(\vartheta^*)+N_R^A(\vartheta^*)}
\end{equation}

\section{Depolarization coefficient }
The depolarization coefficient $D_{NN}$ describes the transfer of the beam-proton polarization to the hyperon. It is positive if the polarization
of the hyperon follows the beam polarization, negative if the hyperon polarization is opposite to the beam polarization and zero, if 
the hyperon polarization is independent of the beam polarization. $D_{NN}$ is given by the equation:

 \begin{equation}\label{eq:DNN}\begin{split}
I(\vartheta^*,\Phi,\theta^*) =&\\I_0(\vartheta^*)\cdot(1&+A_N(\vartheta^*)P_B \cos(\Phi) \\
                                                    &+P_N(\vartheta^*) \:\alpha \cos(\theta^*)\\
                                                    &+D_{NN}(\vartheta^*) \: \alpha P_B \cos(\Phi) \cos(\theta^*))
\end{split}\end{equation}

A formula of the intensity distribution dependent on the angles $\theta^*$ and $\Phi$ is given  for the 
reaction $\bar{p}\vv{p}$$\rightarrow$$\Lambda$$\bar{\Lambda}$ \cite {Paschke2001}. Formula \ref{eq:DNN} is derived from this formula by omitting
the terms, which are related to the $\bar{\Lambda}$ and exchanging the notation for the target polarization with the notation for the 
beam polarization. The  term $D_{NN}(\vartheta^*) \: \alpha P_B \cos(\Phi) \cos(\theta^*)  $
describes the polarization transfer from the beam-proton to the
hyperon. This has to be nearly zero, if the normal of the production plane
is almost perpendicular to the beam-proton polarization direction.
The transfer is maximal if the normal to the production plane has the same
direction as the beam-proton polarization. This dependency is described
with the term $\cos(\Phi)$.

Eight count rates can be defined in order to determine $D_{NN}$. These are extensions to the count rates defined for the hyperon polarization by
adding the two possible beam polarization states:

\noindent

 $\varphi$ in the left hemisphere and the decay particle above the production plane and the beam polarization upwards:

\begin{equation}\label{lau}
N_L^{A \uparrow}(\vartheta^*) = \int_0^1\left[\int_0^1 I\left(\vartheta^*,\Phi,\theta^*\right) \, \mathrm{d}\!\cos\left(\theta^*\right)\right]\, \mathrm{d}\!\cos\left(\Phi\right) 
\end{equation}

 $\varphi$ in the left hemisphere and the decay particle above the production plane and the beam polarization downwards:
\begin{equation}\label{lad}
N_L^{A \downarrow}(\vartheta^*) = \int_{-1}^0\left[\int_0^1 I\left(\vartheta^*,\Phi,\theta^*\right) \, \mathrm{d}\!\cos\left(\theta^*\right)\right]\, \mathrm{d}\!\cos\left(\Phi\right) 
\end{equation}

$\varphi$ in the right hemisphere and the decay particle above the production plane and the beam polarization upwards:

\begin{equation}\label{rau}
N_R^{A \uparrow}(\vartheta^*) = \int_{-1}^0\left[\int_0^1 I\left(\vartheta^*,\Phi,\theta^*\right) \, \mathrm{d}\!\cos\left(\theta^*\right)\right]\, \mathrm{d}\!\cos\left(\Phi\right) 
\end{equation}

$\varphi$ in the right hemisphere and the decay particle above the production plane and the beam polarization downwards:

\begin{equation}\label{rad}
N_R^{A \downarrow}(\vartheta^*) = \int_0^1\left[\int_0^1 I\left(\vartheta^*,\Phi,\theta^*\right) \, \mathrm{d}\!\cos\left(\theta^*\right)\right]\, \mathrm{d}\!\cos\left(\Phi\right) 
\end{equation}

$\varphi$ in the left hemisphere and the decay particle below the production plane and the beam polarization upwards:

\begin{equation}\label{lbu}
N_L^{B \uparrow}(\vartheta^*) = \int_0^1\left[\int_{-1}^0 I\left(\vartheta^*,\Phi,\theta^*\right) \, \mathrm{d}\!\cos\left(\theta^*\right)\right]\, \mathrm{d}\!\cos\left(\Phi\right) 
\end{equation}

 $\varphi$ in the left hemisphere and the decay particle below the production plane and the beam polarization downwards:

\begin{equation}\label{lbd}
N_L^{B \downarrow}(\vartheta^*) = \int_{-1}^0\left[\int_{-1}^0 I\left(\vartheta^*,\Phi,\theta^*\right) \, \mathrm{d}\!\cos\left(\theta^*\right)\right]\, \mathrm{d}\!\cos\left(\Phi\right) 
\end{equation}

$\varphi$ in the right hemisphere and the decay particle below the production plane and the beam polarization upwards:

\begin{equation}\label{rbu}
N_R^{B \uparrow}(\vartheta^*) = \int_{-1}^0\left[\int_{-1}^0 I\left(\vartheta^*,\Phi,\theta^*\right) \, \mathrm{d}\!\cos\left(\theta^*\right)\right]\, \mathrm{d}\!\cos\left(\Phi\right) 
\end{equation}

$\varphi$ in the right hemisphere and the decay particle below the production plane and the beam polarization downwards:

\begin{equation}\label{rbd}
N_R^{B \downarrow}(\vartheta^*) = \int_0^1\left[\int_{-1}^0 I\left(\vartheta^*,\Phi,\theta^*\right) \, \mathrm{d}\!\cos\left(\theta^*\right)\right]\, \mathrm{d}\!\cos\left(\Phi\right) 
\end{equation}
As the asymmetry produced with an up polarized beam in the left hemisphere
is the same as in the right hemisphere produced with a down polarized beam 
(for a symmetrical acceptance and same beam polarization in both directions)
the following count rates are pairwise identically :

$N_L^{A \uparrow}$,$N_R^{A \downarrow}$ and $N_L^{B \downarrow}$,$N_R^{B \uparrow}$ and $N_R^{A \uparrow}$,$N_L^{A \downarrow}$ and $N_R^{B \downarrow}$,$N_L^{B \uparrow}$.
For the following terms the  hyperon polarization and the beam polarization have the same directions:

 \begin{equation*}
N_{same}: N_L^{A \uparrow}+N_R^{B \uparrow}+N_R^{A \downarrow}+N_L^{B \downarrow}
 \end{equation*}

For the following terms the  hyperon polarization and the beam polarization have opposite directions:

 \begin{equation*}
N_{opposite}: N_L^{A \downarrow}+N_R^{B \downarrow}+N_R^{A \uparrow}+N_L^{B \uparrow}
 \end{equation*}

The evaluation of the integrals in equations \ref {lau} - \ref {rbd} yields:

\begin{equation}\label{eq:dnnsame}
N_{same}(\vartheta^*)=I_0(\vartheta^*) (4 + D_{NN}(\vartheta^*) \alpha P_B ) 
\end{equation}

\begin{equation}\label{eq:dnnopposite}
N_{opposite}(\vartheta^*)=I_0(\vartheta^*) (4 - D_{NN}(\vartheta^*) \alpha P_B )
\end{equation}

with the definition of the normalized rate differences as 

\begin{equation}\label{eq:dnn2}
 \epsilon_D (\vartheta^*) = \frac{N_{same}(\vartheta^*)-N_{opposite}(\vartheta^*)}{N_{same}(\vartheta^*)+N_{opposite}(\vartheta^*)}
\end{equation}

the depolarization coefficient can be evaluated:

\begin{equation}\label{eq:dnn5}
D_{NN}(\vartheta^*) = \frac{4} {\alpha P_B} \epsilon_D (\vartheta^*)
\end{equation}

The statistical error of the coefficient is given by
\begin{equation}\label{staterror}
\Delta D_{NN}(\vartheta^*) = \frac{8} {\alpha P_B}  \sqrt {  \frac{N_{same}(\vartheta^*)N_{opposite}(\vartheta^*)}
                                                                 {(N_{same}(\vartheta^*)+N_{opposite}(\vartheta^*))^3}}
\end{equation}

\subsection{Application to  measurements with limited acceptance }
For experiments with limited acceptance the relevant angular regions have to be taken into account for the integration limits.
 In the Brookhaven Multiparticle Spectrometer  \cite{Bonner1988} for example only particles are detected, which are produced on the 
left side of the proton beam. In addition, the acceptance restricts the production plane to be nearly horizontal, this
allows in the derivation of the $D_{NN}$ formula  to set $\Phi$ = 0 or $\pi$ (in equation 12 of \cite{Bonner1988}).
Regarding only the count rate of equations \ref{lau} - \ref{rbd} in the left hemisphere and setting $\Phi$ = 0 for beam polarization
upwards and $\Phi$ = $\pi$ for beam polarization downwards the following equations are obtained:

\begin{equation}\label{b-lau}
N_L^{A \uparrow}(\vartheta^*) = 1 + P_B A_N + \frac{1}{2}P_N \alpha + \frac{1}{2} P_B \alpha D_{NN} 
\end{equation}

\begin{equation}\label{b-lad}
N_L^{A \downarrow}(\vartheta^*) = 1 - P_B A_N + \frac{1}{2}P_N \alpha - \frac{1}{2} P_B \alpha D_{NN} 
\end{equation}

\begin{equation}\label{b-lbu}
N_L^{B \uparrow}(\vartheta^*) = 1 + P_B A_N - \frac{1}{2}P_N \alpha - \frac{1}{2} P_B \alpha D_{NN} 
\end{equation}

\begin{equation}\label{b-lbd}
N_L^{B \downarrow}(\vartheta^*) = 1 - P_B A_N - \frac{1}{2}P_N \alpha + \frac{1}{2} P_B \alpha D_{NN} 
\end{equation}


Inserting these count rates into equations \ref {eq:hyperonpolarization2} and \ref {eq:hyperonpolarization3}
the following relation is obtained:

\begin{equation*}
P^{\uparrow\downarrow}= \frac{P_N \pm P_B D_{NN}}{1 \pm P_B A_N}
\end{equation*}

which corresponds to equation 13 in \cite{Bonner1988}.

Applying the same conditions of  $\Phi$ on the equation \ref{eq:DNN} yields:

\begin{equation}\label{eq:b-DNN1}\begin{split}
I(\vartheta^*,\Phi=0,\theta^*)^{\uparrow} =&\\I_0(\vartheta^*)\cdot(1&+A_N(\vartheta^*)P_B 
                                          \\ &+P_N(\vartheta^*) \: \alpha \cos(\theta^*)
                                          \\ &+D_{NN}(\vartheta^*) \: \alpha P_B \cos(\theta^*))
\end{split}\end{equation}

\begin{equation}\label{eq:b-DNN2}\begin{split}
I(\vartheta^*,\Phi=0,\theta^*)^{\downarrow} =&\\I_0(\vartheta^*)\cdot(1&-A_N(\vartheta^*)P_B 
                                          \\ &+P_N(\vartheta^*)  \:\alpha \cos(\theta^*)
                                          \\ &-D_{NN}(\vartheta^*)  \:\alpha P_B \cos(\theta^*))
\end{split}\end{equation}

The asymmetry of the intensities in equations \ref{eq:b-DNN1} and \ref{eq:b-DNN2} is:

\begin{equation}
\frac{I(\theta^*)^{\uparrow}-I(\theta^*)^{\downarrow}}{I(\theta^*)^{\uparrow}+I(\theta^*)^{\downarrow}}
= \frac{P_B A_N+P_B D_{NN} \alpha  \cos(\theta^*)}{1+\alpha P_N \cos(\theta^*) }
\end{equation}

this corresponds to equations 15 and 16 in \cite{Bonner1988}.

\section{Conclusion}
For the associated strangeness reactions  $\vv{\mathrm{p}}$p $\rightarrow$ p$\mathrm{K}^{+}\Lambda$ 
and $\vv{\mathrm{p}} $p $\rightarrow$ p$\mathrm{K}^{0}\Sigma^{+}$  it is shown that the depolarization
coefficient according to the ``integral method'' is given by 

\begin{equation*}\label{eq:dnn}
D_{NN}(\vartheta^*) = \frac{4} {\alpha P_B} \epsilon_D (\vartheta^*)
\end{equation*}

The asymmetry $\epsilon_D (\vartheta^*)$ can be expressed by 8 different count rates depending on the beam
polarization, the position in the left or right hemisphere, and the position above or below the production 
plane. 
Applying restrictions in the acceptance of other measurements as for instance of the Brookhaven Multiparticle Spectrometer  \cite{Bonner1988} 
onto this formula, the corresponding formulas  are derived. 
\newline
\newline
\section*{Acknowledgments}
This work has been supported by COSY-FFE. We acknowledge valuable discussions with J. Haidenbauer.


\bibliographystyle{epj} 
\bibliography{dnn}

\end{document}